# Zipf-Mandelbrot Law for Time Intervals of Earthquakes

Sumiyoshi Abe[1] and Norikazu Suzuki[2]

[1] *Institute of Physics, University of Tsukuba, Ibaraki 305-8571, Japan*

[2] *College of Science and Technology, Nihon University, Chiba 274-8501, Japan*

The statistical properties of the seismic time series data in southern California are studied. In particular, the calm time intervals, which are the time intervals between successive significant earthquakes above the fixed threshold value of magnitude, are analyzed. It is found that the system undergoes a series of transition between stationary states, whose cumulative waiting time distributions all obey the Zipf-Mandelbrot law, exhibiting a new scale free nature of earthquakes. Dependencies of the two parameters characterizing the Zipf-Mandelbrot distribution on the threshold are also discussed.

Although seismicity is characterized by extraordinarily rich phenomenology and complexity, which make it difficult to develop coherent explanation and prediction of earthquakes, some empirical laws are known to hold, which are remarkably simple. Classical examples are the Omori law [1] for temporal pattern for aftershocks and the Gutenberg-Richter law [2] for frequency and magnitude. These are power laws and represent scale free natures of earthquake phenomena. Also, it should be mentioned that quite recently an attempt has been made to unify the Omori law and the Gutenberg-Richter law within a single hypothetical scaling law [3].

One of the extreme goals of seismology is to predict when the next main shock will come after some earlier main shock. Though science seems still far from this goal, there exist some interesting approaches based on peculiar properties of precursory phenomena (see Refs. [4-6] and the references threrein). As a step toward prediction, it may be of obvious importance to fully elucidate the statistical properties of the observed complex random seismic time series.

Recently, we have studied the Internet time series obtained by performing the Ping experiments [7-9]. We have found that there exists a striking similarity between the Internet and seismic time series. Like the seismic ones, the Internet time series can also be characterized by two different time scales: one is the short time scale associated with sudden heavy congestion of the Internet traffic, which is regarded as the Internetquake, and the other is a long one corresponding to user's time scale. "Magnitude" of the Internetquake is defined with respect to the round-trip time of the Ping signal [9]. We have shown that the Omori law [8] and the Gutenberg-Richter law [9] also holds for the Internetquakes. Furthermore, we have studied the waiting time distribution of the Internet time series [7], which is the distribution of the sparseness time intervals between two successive significant congested states. We have found that the observed distributions are well described by the family of the $q$-exponential distributions, which can be thought of as a generalization of the Zipf-Mandelbrot distribution [10] (see below). The network is observed to itinerate over a series of the stationary states characterized by these generalized Zipf-Mandelbrot distributions. Therefore, a corresponding natural question occurring in seismology is: what are the statistical properties of the time intervals between two successive earthquakes?

In this Letter, we report the discovery of a new scale free nature in earthquake phenomena. We analyze a set of the observed data of magnitude of earthquakes in southern California. We set up a threshold value of magnitude in an appropriate manner and then measure the time interval between two successive values of magnitude above a threshold value, which is referred to here as the "calm time interval", $\tau$. A collection of the calm time intervals thus obtained gives a histogram, from which the cumulative waiting time distribution, $P(>\tau)$, can be calculated. We show that the time series thus obtained for the calm time interval may be composed of a series of stationary states, each of which is always described by the distribution of the Zipf-Mandelbrot type, exhibiting a scale free nature of such stationary states. We also discuss dependencies of the two parameters characterizing the distribution on the threshold.

We have analyzed the earthquake catalog made available by the Southern California Earthquake Data Center (http://www.scecdc.scec.org/catalogs.html) covering the period

between 08:03:42 on 19 October, 1999 and 9:40:50 on 1 November, 1999 in the region spanning $31^\circ 12' 71'' \mathrm{N} - 38^\circ 08' 64'' \mathrm{N}$ latitude and $114^\circ 37' 06'' \mathrm{W} - 121^\circ 28' 05'' \mathrm{W}$ longitude.

In Fig. 1, we present the time series of the calm time interval with the threshold value of magnitude $m_{\mathrm{th}} = 2.5$. There does not seem to exist a simple operational way of identifying the stationary regimes, in general, as long as based only on this figure. However, according to our preliminary examinations, it seems plausible to identify the stationary regimes by considering the "intervals of the calm time intervals", which could be regarded as a kind of the differential method. We wish to discuss the details of this problem elsewhere. In the present work, we identify the three regimes: a, b, and c.

In Figs. 2a-c, we present the log-log plots of the cumulative distributions associated with the statistical frequencies of the calm time interval. The dots represent the observed cumulative distributions, whereas the solid lines are of the power-law distributions of the Zipf-Mandelbrot type, which may be expressed by [10]

$$P(>\tau) = \frac{1}{(1+\varepsilon\tau)^{\alpha}} \qquad (\alpha, \varepsilon > 0). \tag{1}$$

For the sake of convenience, we rewrite this in the following form:

$$P(>\tau) = e_q(-\tau / \tau_0), \tag{2}$$

where $e_q(x)$ stands for the $q$-exponential function defined by

$$
e_q(x) = \begin{cases} [1+(1-q)x]^{1/(1-q)} & (1+(1-q)x \geq 0) \\ 0 & (1+(1-q)x < 0) \end{cases} . \tag{3}
$$

In this parametrization, $q$ and $\tau_0$ are obviously related to $\alpha$ and $\varepsilon$ in Eq. (1) as $\alpha = 1/(q-1)$ and $\varepsilon = (q-1)/\tau_0$. The inverse function of the $q$-exponential function is the $q$-logarithmic function given by

$$
\ln_q(x) = \frac{x^{1-q}-1}{1-q} . \tag{4}
$$

In the limit $q \to 1$, $e_q(x)$ and $\ln_q(x)$ tend to the ordinary exponential and logarithmic functions, respectively. The Zipf-Mandelbrot distribution is a scale-invariant power-law distribution with a long tail, corresponding to the exponent $q > 1$. In fact, so far, only the case $q > 1$ has been observed in our extensive examinations. To show how the observations are well described by the distributions of the Zipf-Mandelbrot type, we also present the plots on "semi-$q$-log scale" in the inside boxes. Thus, we arrive at the main result that the system of earthquakes in view of the time series of the calm time interval undergoes a series of transition from one stationary state to another: $(q_1, \tau_{0,1}) \to (q_2, \tau_{0,2}) \to (q_3, \tau_{0,3}) \to \cdots$, and each stationary state is described by the scale-invariant distribution of the Zipf-Mandelbrot type.

Finally, we wish to make a comment on dependencies of the parameters of the distribution on the threshold, $m_{\text{th}}$. Let us examine this by taking the regime a. In Fig. 3, we present the plot of $q$ and $\tau_0$ for $m_{\mathrm{th}}$ ranging from 0 to 2.5. One sees that both $q$ and $\tau_0$ monotonically increase with respect to $m_{\mathrm{th}}$. Through our extensive examinations using other regimes, we have ascertained that this tendency is in fact universal.

In conclusion, we have discovered a new scale free nature for the calm time intervals between significant earthquakes. We have observed that the complex system of earthquakes undergoes a series of stationary states, which are all described by the power-law distributions of the Zipf-Mandelbrot type. It is worth mentioning that there exists a statistical mechanical theory for such distributions. The Tsallis entropy indexed by $q$ [11,12] is known to be maximized by them under appropriate constraints on the averages of the physical quantities (e.g., the calm time interval) to be measured. This theory is now expected to describe complex systems at their stationary states [13-15]. An important point arising here is that this fact enables us to develop a thermodynamic approach to seismology and should provide new insights into the study of earthquakes.

S. A. is supported in part by the internal research project of the University of Tsukuba.

# Figure Captions

Fig. 1  Time series data of the calm time interval. The origin is adjusted relatively to 00:00:00 on 1 January, 1999. The total number of events is 19889 and three stationary regimes, a, b, and c, are identified.

Fig. 2  a: Between 08:03:42 on 19 October, 1999 and 09:40:50 on 09:40:50 on 1 November, 1999. The number of events is 3064. $q = 1.37$ and $\tau_0 = 1.20 \times 10^3$ $s$. The straight line on the semi-$q$-log scale corresponds to the value of correlation coefficient $\rho = -0.993$. b: Between 09:40:50 on 09:40:50 on 1 November, 1999 and 13:21:56 on 18 December, 1999. The number of events is 5580. $q = 1.20$ and $\tau_0 = 8.00 \times 10^3$ $s$. The straight line on the semi-$q$-log scale corresponds to the value of correlation coefficient $\rho = -0.993$. c: Between 13:21:56 on 18 December, 1999 and 13:21:56 on 18 December, 1999 and 11:36:14 on 2 April, 2000. The number of events is 6695. $q = 1.07$ and $\tau_0 = 2.04 \times 10^4$ $s$. The straight line on the semi-$q$-log scale corresponds to the value of correlation coefficient $\rho = -0.996$. The values of correlation coefficient are obtained by making use of the method of least squares.

Fig. 3  Dependencies of $q$ (dimensionless) and $\tau_0$ on the threshold in the regime a in Fig. 1. From the bottom to the top, $m_{\text{th}} =$ 0.5, 1.5, 1.6, 1.7, 1.8, 1.9, 2.0, 2.1, 2.2, 2.3, 2.4, 2.5.

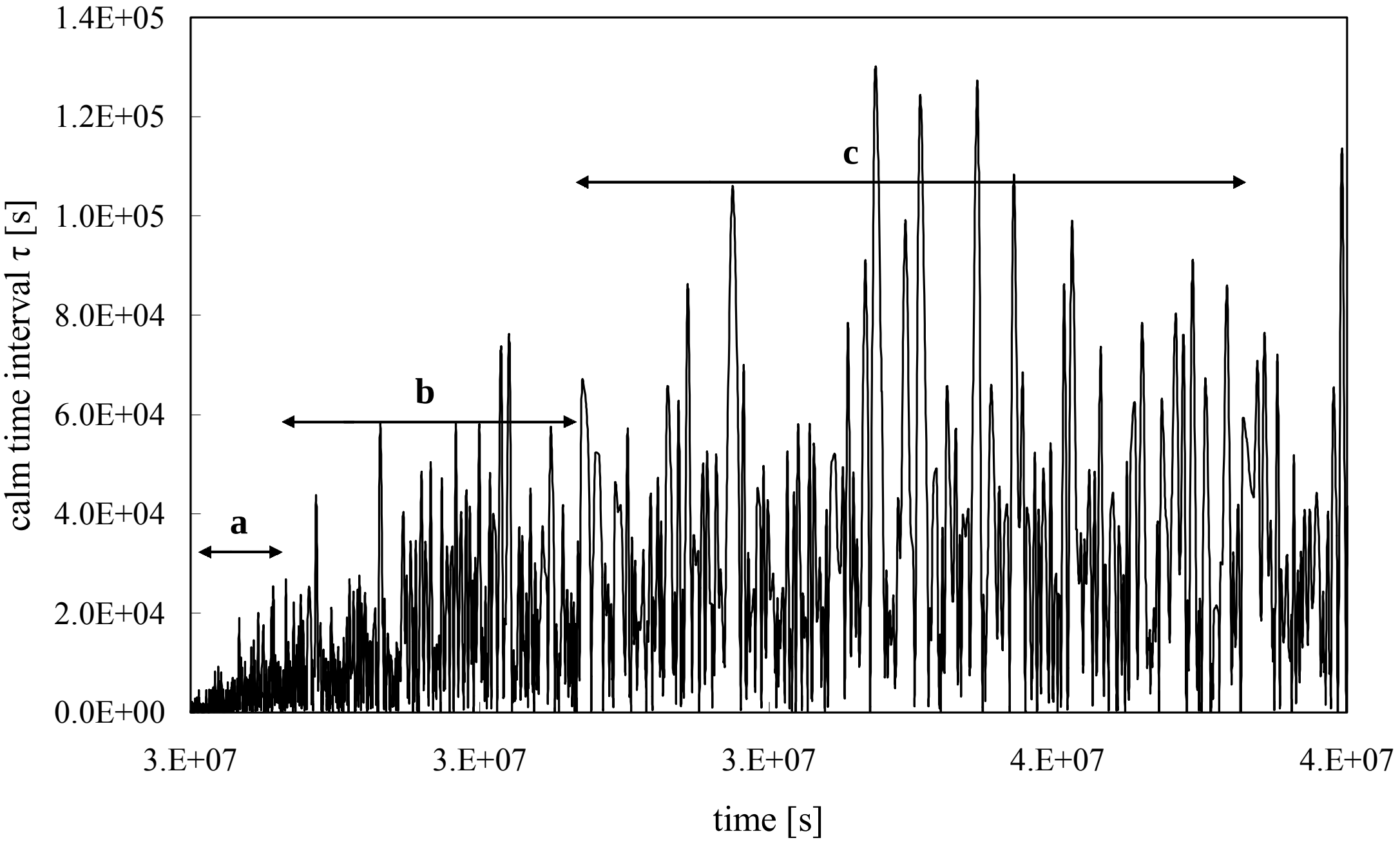

**Fig. 1**

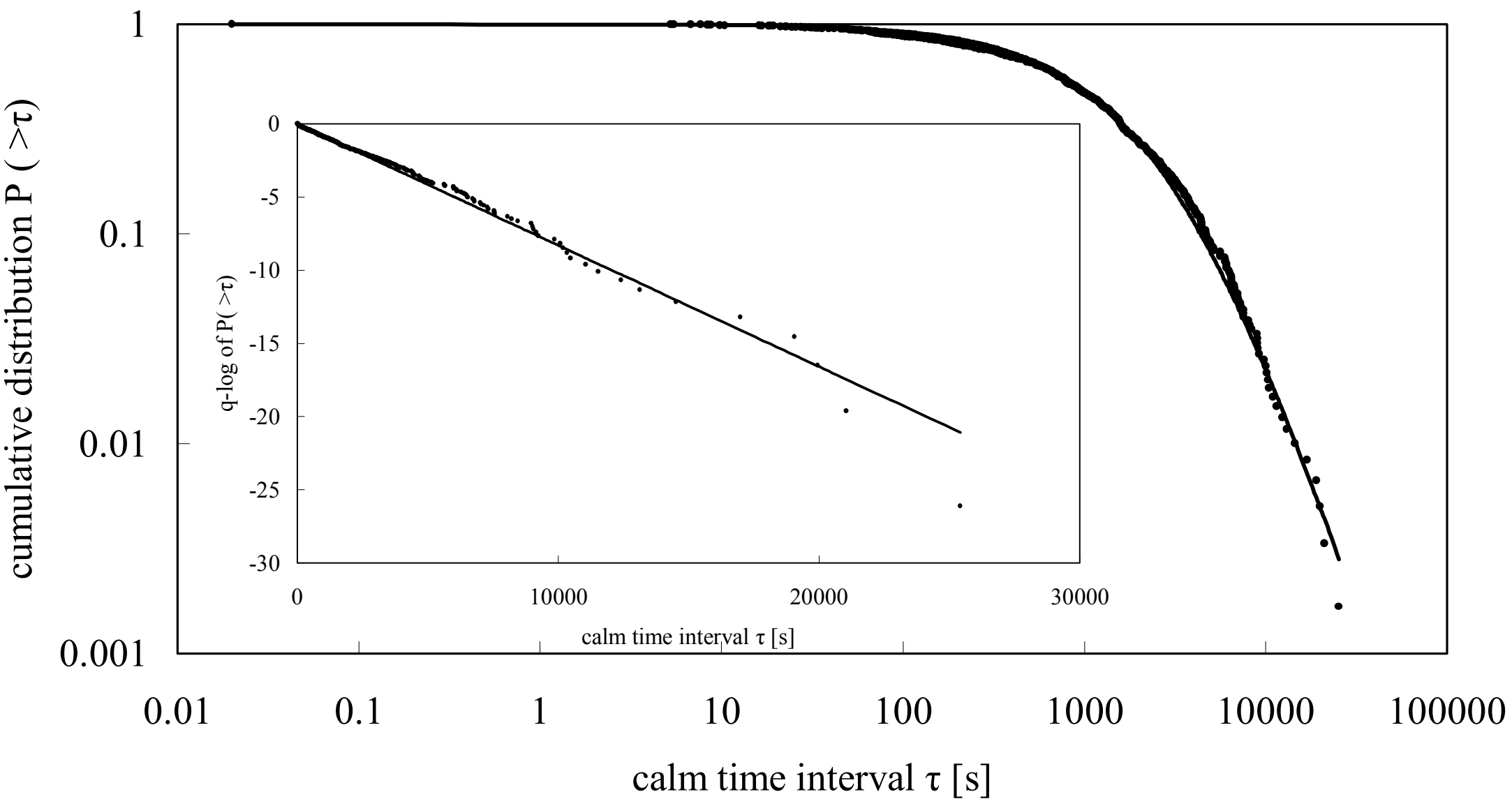

**Fig. 2a**

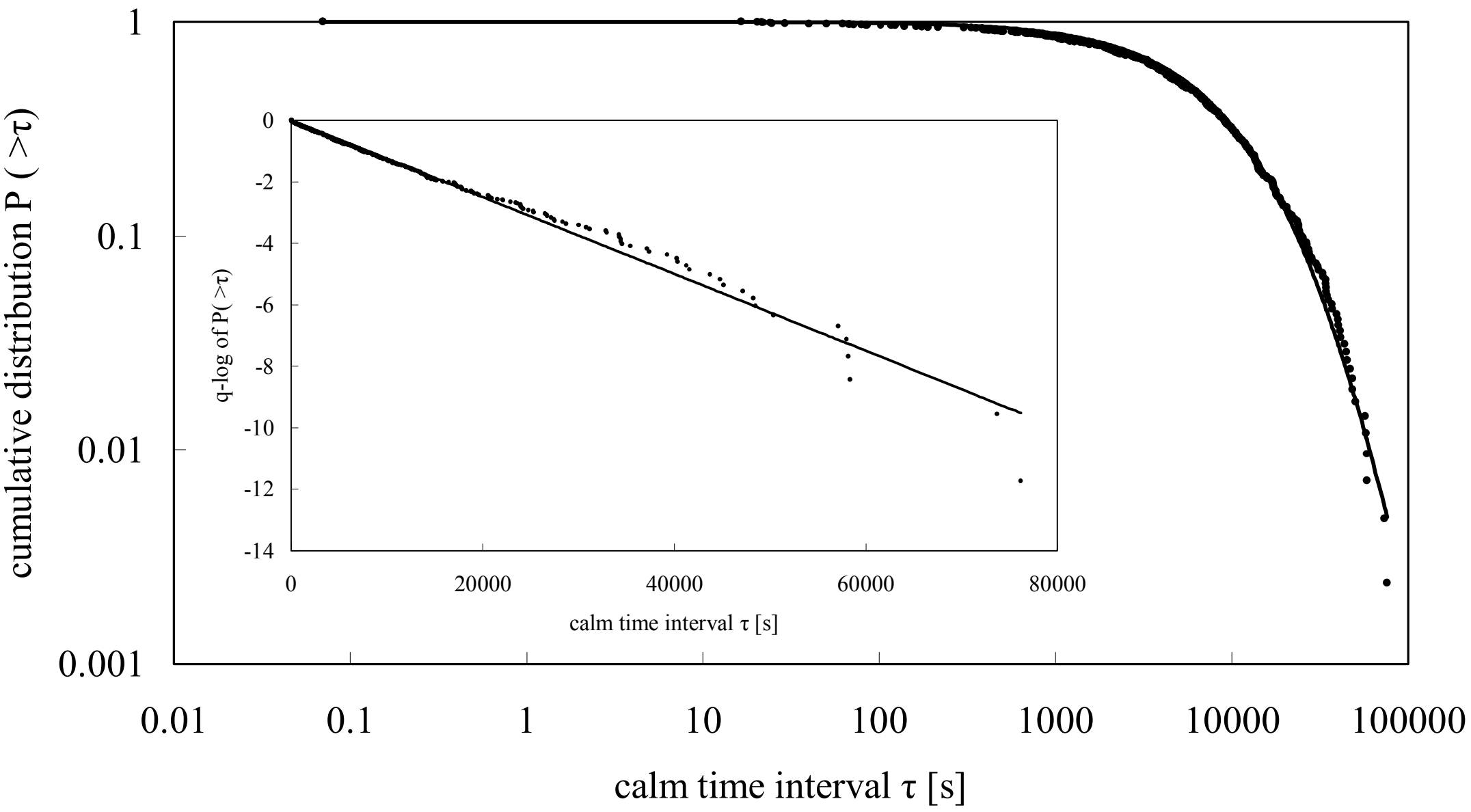

**Fig. 2b**

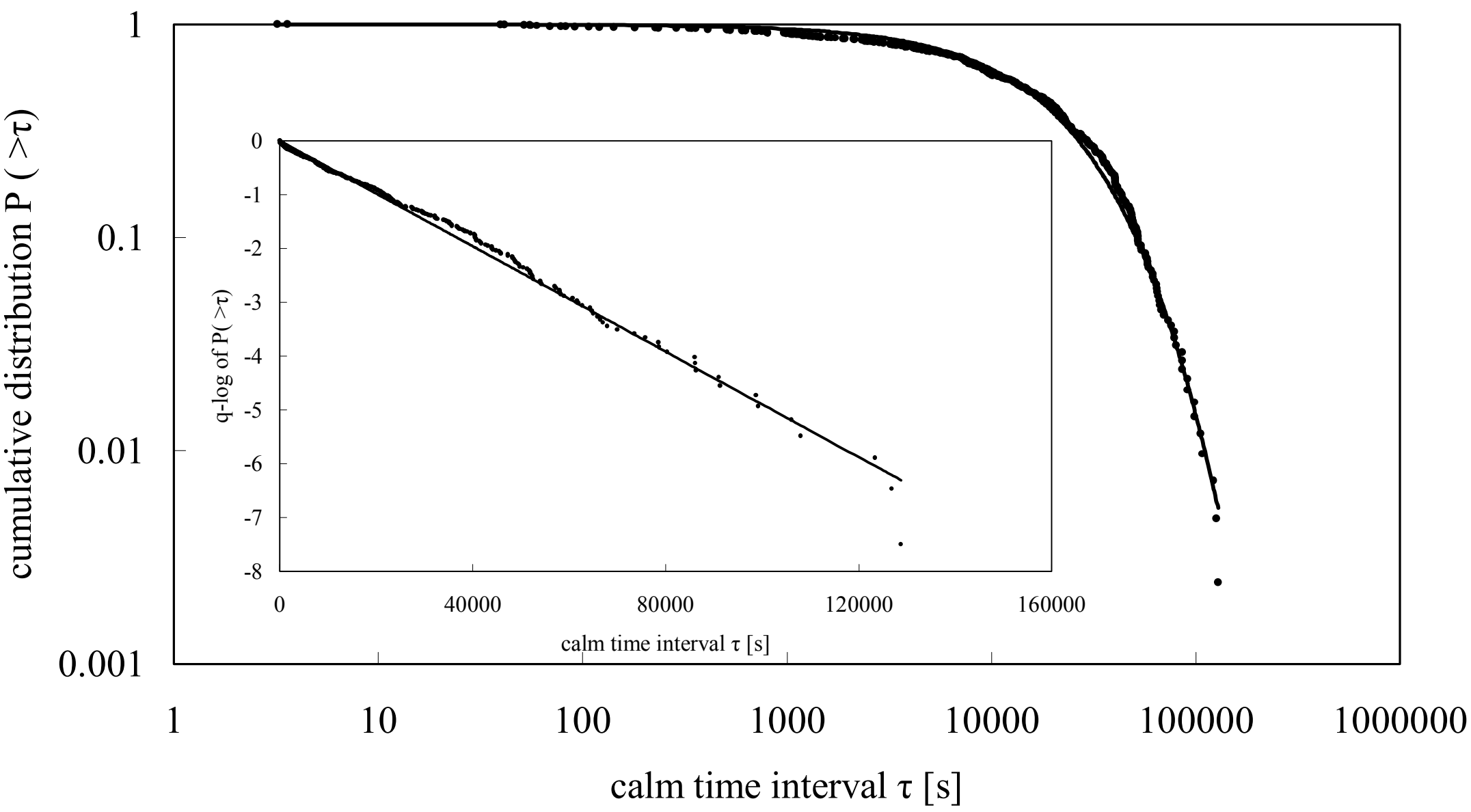

**Fig. 2c**

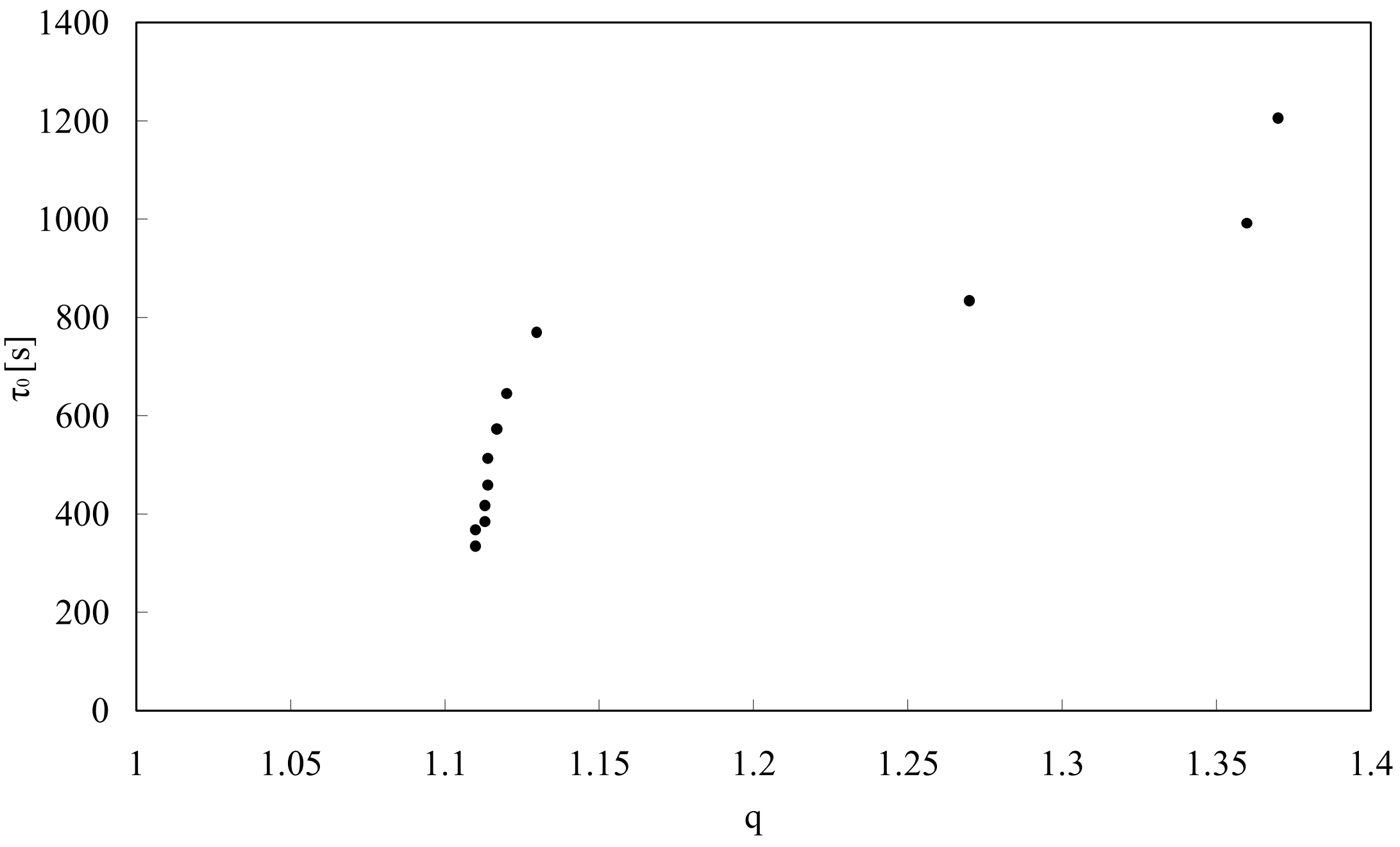

**Fig. 3**